\documentclass[12pt]{iopart}
\eqnobysec

\begin{document}
 \jl{1}
\title{Resonance photon generation in a vibrating cavity}
\author{V V Dodonov\footnote{On leave from Lebedev Physical Institute
and Moscow Institute of Physics and Technology, Russia}
\footnote{E-mail: vdodonov@power.ufscar.br}}
\address{
Departamento de F\'{\i}sica, Universidade Federal de S\~ao Carlos,\\
Via Washington Luiz km 235, 13565-905  S\~ao Carlos,  SP, Brazil}

 \date{\today}

\begin{abstract}
The problem of photon creation from vacuum due to the nonstationary
Casimir effect in an ideal one-dimensional
Fabry--Perot cavity with vibrating walls is solved in the resonance case,
when the frequency of vibrations is close to the frequency of
some unperturbed electromagnetic mode:
$\omega_w=p(\pi c/L_0)(1+\delta)$, $|\delta|\ll 1$, $p=1,2,\ldots$
($L_0$ is the mean distance between the walls).
An explicit analytical expression for the total energy in all the modes
shows an exponential growth if $|\delta|$ is less than the
dimensionless amplitude of vibrations $\varepsilon\ll 1$,
the increment being proportional to $p\sqrt{\varepsilon^2-\delta^2}$.
The rate of photon generation from vacuum in the $(j+ps)$th mode goes
asymptotically to a constant value
$cp^2\sin^2(\pi j/p)\sqrt{\varepsilon^2-\delta^2}
/[\pi L_0 (j+ps)]$,
the numbers of photons in the modes with indices $p,2p,3p,\ldots$
being the integrals of motion.
The total number of photons in all the modes is proportional to
$p^3(\varepsilon^2-\delta^2) t^2$ in the short-time and in the long-time
limits.
In the case of strong detuning $|\delta|>\varepsilon$
the total energy and the total number of photons generated from vacuum
oscillate with the amplitudes decreasing as
$(\varepsilon/\delta)^2$ for $\varepsilon\ll|\delta|$.
The special cases of $p=1$ and $p=2$ are studied in detail.
\end{abstract}

\submitted
\maketitle

\section{Introduction}

Fifty years ago Casimir \cite{Cas} showed that the presence of boundaries
changes the ground state of the electromagnetic field, leading to nontrivial
quantum effects like the {\it Casimir force\/} (see also [2-4]).
%\cite{plun1,Mil,Most}).
Last years, an attention of many authors [5-40]  was attracted to the
{\it nonstationary modifications\/} of the Casimir effect in the
case of {\it moving boundaries\/}
(a detailed list of publications
before 1995 was given in \cite{DKPR}).
The present article is devoted to the special case of the {\it nonstationary
Casimir effect\/} (NSCE), namely, to the effect of
{\it photon creation from vacuum\/} in an ideal one-dimensional cavity
(a model of the Fabry--Perot interferometer) with {\it vibrating\/}
boundaries.

As was understood recently [10,14-23],
%\cite{DK92,DKN93,Law,LawPRL,Law-new,D95,Cole,Mep,DKPR,D96,Lamb},
notwithstanding that the maximal velocity of the boundary achievable
under the laboratory conditions is very small in comparison with the
speed of light,
a gradual accumulation of the small changes in the quantum state of the
field could result finally in a significant observable effect,
if the boundaries of a cavity perform
small oscillations at a frequency $\omega_w$ which is an integer multiple
of the unperturbed eigenfrequency of the fundamental electromagnetic mode
$\omega_1=\pi c/L_0$
(where $L_0$ is the mean distance between the walls):
$\omega_w=p\omega_1$, $p=1,2,\ldots$ (remember that the spectrum of the
electromagnetic modes is equidistant in the case involved: the unperturbed
frequency of the $p$th mode equals $\omega_p=p\omega_1$).
The time evolution of the field in the short time limit
$\varepsilon\omega_1 t\ll 1$ (where $\varepsilon\ll 1$ is a ratio
of the amplitude of vibrations to $L_0$) was considered in
\cite{Sarkar,DKM90} in the framework of Moore's approach
\cite{Moore} and in \cite{Calucci,Law,Ji,Plun,Ji98} in the framework of
the ``instantaneous basis'' method (IBM) described in section 2.
The asymptotical solutions to Moore's equation
in the case $\varepsilon\omega_1 t\gg 1$ were obtained
in \cite{DK92,DKN93,Klim}, and more general solutions were found in
\cite{LawPRL,Cole,Dal}.
A detailed study of the problem in the framework of the
IBM was given in \cite{DKPR} for $p=2$ and in \cite{D96} for $p=1$.
The {\it short-time\/} limit $\varepsilon\omega_1 t\ll 1$ for an arbitrary
integer value of $p$ was considered in \cite{Ji}.
However, in all the cited papers the solutions were found under the
condition of the {\it strict resonance\/} $\omega_w=\omega_p$
between the mechanical and
electromagnetic oscillations (excepting recent
article \cite{D98}, where a detuned three-dimensional cavity with a
nondegenerate spectrum was considered). Evidently, such a condition is
an idealization.

The aim of the present paper is to study the case of a
{\it nonzero\/} (although small) detuning between the frequencies of
the mechanical and field modes:
\begin{equation}
\omega_w=p\omega_1(1+\delta),
\quad |\delta|\ll 1
\label{omegw}
\end{equation}
for any integer $p=1,2,\ldots$, thus generalizing the results of
\cite{DKPR,D96,Ji}.
It will be shown that the photons can be created
from vacuum provided the dimensionless detuning parameter does not exceed the
dimensionless amplitude of the wall vibrations, otherwise the total number
of photons generated inside the cavity exhibits small oscillations and goes
periodically to zero.

The plan of the paper is as follows. In section~2 we give general formulae
related to the field quantization in a cavity with moving boundaries
and derive the simplified ``reduced equations'' in the resonance case.
A simple explicit analytical expression for
the total energy of the field in all the modes is found in section~3.
Section~4 is devoted to the ``semi-resonance'' case $p=1$
when the frequency of the wall is close to the
fundamental frequency of the field. Under this condition
new photons are not created, but the total energy of all the field modes
increases exponentially with time above the threshold or oscillates in the
case of a large detuning.
The generic resonance case of an arbitrary $p\ge2$ is analysed in section~5
and the simplifications in the case $p=2$ are considered in section~6.
A brief discussion of the results is given in section~7.
Some details of calculations are given in the appendix.

\section{Field quantization and reduced equations in the resonance case}

Following the scheme of the field quantization in a cavity with
time-dependent boundary conditions first proposed by Moore \cite{Moore},
we consider a cavity formed by two infinite ideal plates moving in
accordance with the prescribed laws
\[
x_{left}(t)=u(t), \quad
x_{right}(t)=u(t)+L(t)
\]
where $L(t)>0$ is the time dependent length of the cavity.
Taking into account only the electromagnetic modes
whose vector potential is directed along $z$-axis (``scalar electrodynamics''),
one can write down
the  field operator {\em in the Heisenberg representation\/}
$\hat {A}(x,t)$ at $t\le 0$ (when both the plates were at rest at the
positions $x_{left}=0$ and $x_{right}=L_0$) as (we assume $c=\hbar=1$)
\begin{equation}
\hat {A}_{in}=2\sum_{n=1}^{\infty}\frac 1{\sqrt {n}}\sin\frac {
n\pi x}{L_0}\hat b_n\exp\left(-i\omega_nt\right)+\mbox{h.c.}
 \label{Ast}
 \end{equation}
where $\hat {b}_n$ means the usual annihilation photon
operator and $\omega_n=\pi n/L_0$.
The choice of coefficients in equation (\ref{Ast})
corresponds to the standard form of the field Hamiltonian
\begin{equation}
\hat {H}\equiv\frac 1{8\pi}\int_0^{L_0}\mbox{d}x\,\left
[\left(_{}\frac {\partial A}{\partial t}\right)^2+\left(_{}\frac {
\partial A}{\partial x}\right)^2\right]
=\sum_{n=1}^{\infty}\omega_n\left(\hat
b^{\dag}_n\hat b_n+\frac 12\right). \label{Ham}
\end{equation}
For $t>0$ the field operator can be written as
\[ \hat {A}(x,t)=2\sum_{n=1}^{\infty}\frac 1{\sqrt {n}}\left[
\hat b_n\psi^{(n)}(x,t)\,+\,\mbox{h.c.}\,\right].\]
To find the explicit form of functions $\psi^{(n)}(x,t)$, $n=1,2,\ldots$,
one should take into account that the field operator must satisfy

i) the wave equation
\begin{equation}
\frac {\partial^2A}{\partial t^2}\,-\frac {\partial^
2A}{\partial x^2}=0,
\label{we}
\end{equation}

ii) the boundary conditions
\begin{equation}
A(u(t),t)=A(u(t)+L(t),t)=0,
\label{boundcon}
\end{equation}

iii) the initial condition (\ref{Ast}), which is equivalent to
\begin{equation}
\psi^{(n)}\left(x,t<0\right)=\sin\frac {n\pi x}{L_0}
\exp\left(-i\omega_nt\right).
\label{init}
\end{equation}
Following the approach of Refs. \cite{Calucci,Law,Law-new}
we expand the function $\psi^{(n)}(x,t)$
in a series with respect to the {\em instantaneous basis\/}:
\begin{equation}\psi^{(n)}(x,t>0)=\sum_{k=1}^{\infty}
Q_k^{(n)}(t)\sqrt {\frac {
L_0}{L(t)}}\sin\left(\frac {\pi k[x-u(t)]}{L(t)}\right),
\quad n=1,2,\ldots
\label{psit}
\end{equation}
with the initial conditions
\[
Q_k^{(n)}(0)=\delta_{kn},\quad\dot {Q}_k^{(n)}(0)=-i\omega_n\delta_{kn},
\quad k,n=1,2,\ldots
\]
This way we satisfy automatically both the boundary conditions
(\ref{boundcon}) and the initial condition (\ref{init}).
Putting expression  (\ref{psit}) into the wave equation (\ref{we}),
one can arrive after some algebra at an infinite set of coupled
differential equations \cite{Plun,Ji98} ($k,n=1,2,\ldots$)
\begin{equation}
\ddot {Q}_k^{(n)}+\omega_k^2(t)Q_k^{(n)}
=2\sum_{j=1}^{\infty}  g_{kj}(t)\dot {Q}_j^{(n)}+
\sum_{j=1}^{\infty}  \dot{g}_{kj}(t) Q_j^{(n)}
+{\cal O}\left(g_{kj}^2\right),
\label{Qeq}
\end{equation}
where
\[
\omega_k(t)= {k\pi}/{L(t)}
\]
and the time dependent antisymmetric coefficients $g_{kj}(t)$ read
($j\neq k$)
\begin{equation}
g_{kj}=-g_{jk}=(-1)^{k-j}\frac {2kj
\left(\dot {L} +\dot {u}\epsilon_{kj}\right)}{\left(j^2-k^2\right)L(t)},
\quad
\epsilon_{kj}= 1-(-1)^{k-j}.
\label{gkj}
\end{equation}
For $u=0$ (the left wall at rest) the equations like
(\ref{Qeq})-(\ref{gkj}) were derived in \cite{Calucci,Law-new}.

If the wall comes back to its initial position $L_0$ after some interval of
time $T$, then the right-hand side of equation (\ref{Qeq}) disappears, so
at $t>T$ one gets
\begin{equation}
Q_k^{(n)}(t)=\xi_k^{(n)}e^{-i\omega_kt}+\eta_k^{(n)}e^{i\omega_kt},
\quad k,n=1,2,\ldots
\label{ksi}
\end{equation}
$\xi_k^{(n)}$ and $\eta_k^{(n)}$ being some constant coefficients.
Consequently, at $t>T$ the initial annihilation operators $\hat {b}_n$
cease to be ``physical'', due to the contribution of the terms with
``incorrect signs'' in the exponentials $\exp(i\omega_kt)$. Introducing
a new set of ``physical'' operators $\hat {a}_m$
and $\hat {a}_m^{\dag}$, which result at $t>T$ in relations such as (\ref{Ast})
and (\ref{Ham}), but with $\hat {a}_m$ instead of $\hat {b}_m$, one can
easily check that the two sets
of operators are related by means of the Bogoliubov transformation
\begin{equation}
\hat {a}_m=\sum_{n=1}^{\infty} \left(\hat b_n\alpha_{nm}+\hat b_
n^{\dag}\beta_{nm}^{*}\right), \quad m=1,2,\ldots
\label{Bogol}
\end{equation}
with the coefficients
\begin{equation}\alpha_{nm}=\sqrt {\frac mn}\xi_m^{(n)},\qquad\beta_{
nm}=\sqrt {\frac mn}\eta_m^{(n)}.\label{al-ksi}
\end{equation}
The unitarity of the transformation (\ref{Bogol}) implies the following
constraints:
\begin{eqnarray}
&&\sum_{m=1}^{\infty}\left(\alpha_{nm}^*\alpha_{km} -
\beta_{nm}^*\beta_{km}\right)
= \sum_{m=1}^{\infty} \frac{m}{n}
\left(\xi_{m}^{(n)*}\xi_{m}^{(k)} -
\eta_{m}^{(n)*}\eta_{m}^{(k)} \right) =\delta_{nk}
\label{cond1} \\[2mm]
&&\sum_{n=1}^{\infty}\left(\alpha_{nm}^*\alpha_{nj} -
\beta_{nm}^*\beta_{nj}\right)
= \sum_{n=1}^{\infty}\frac{m}{n}
\left(\xi_{m}^{(n)*}\xi_{j}^{(n)} -
\eta_{m}^{(n)*}\eta_{j}^{(n)} \right) =\delta_{mj}
\label{cond2}\\[2mm]
&&\sum_{n=1}^{\infty}\left(\beta_{nm}^*\alpha_{nk} -
\beta_{nk}^*\alpha_{nm}\right)
= \sum_{n=1}^{\infty}\frac{1}{n}
\left(\eta_{m}^{(n)*}\xi_{k}^{(n)} -
\eta_{k}^{(n)*}\xi_{m}^{(n)} \right) =0
\label{cond3}
\end{eqnarray}

The mean number of photons in the $m$th mode equals the
average value of the operator $\hat {a}_m^{\dag}\hat {a}_m$ in the initial
state $|{\rm in}\rangle$ (remember that we use the Heisenberg picture),
since just this operator has a physical meaning at $t>T$:
\begin{eqnarray}
&&{\cal N}_m \equiv \langle {\rm in}|\hat {a}_m^{\dag}
\hat {a}_m|{\rm in}\rangle \nonumber\\[2mm]
&&=\sum_n|\beta_{nm}|^2 +\sum_{n,k}\left[\left(\alpha_{nm}^*\alpha_{km}
+\beta_{nm}^*\beta_{km}\right)\langle\hat {b}_n^{\dag}\hat {b}_k\rangle
+ 2 {\rm Re}\left(\beta_{nm}\alpha_{km}\langle\hat {b}_n\hat {b}_k
\rangle\right)\right]
\nonumber\\[2mm]
&&= \sum_{n=1}^{\infty}\frac{m}{n}|\eta_m^{(n)}|^2
+ \sum_{n,k=1}^{\infty} \frac{m}{\sqrt{nk}}
\left(\xi_{m}^{(n)*}\xi_{m}^{(k)} +
\eta_{m}^{(n)*}\eta_{m}^{(k)} \right)
\langle\hat {b}_n^{\dag}\hat {b}_k\rangle
\nonumber\\[2mm]
&&+ 2{\rm Re}\sum_{n,k=1}^{\infty} \frac{m}{\sqrt{nk}}
\eta_{m}^{(n)}\xi_{m}^{(k)}\langle\hat {b}_n\hat {b}_k \rangle .
\label{number}
\end{eqnarray}
The first sum in the right-hand sides of the relations above describes the
effect of the photon creation from vacuum due to the NSCE,
while the other sums are different from zero only in the case of a
nonvacuum initial state of the field.

To find the coefficients $\xi_k^{(n)}$ and $\eta_k^{(n)}$ one has to solve
an infinite set of coupled equations (\ref{Qeq}) ($k=1,2,\ldots$)
with time-dependent coefficients, moreover, each
equation also contains an infinite number of terms.
However, the problem can be essentially simplified, if the walls
perform small oscillations at the frequency $\omega_w$ close to some
unperturbed field eigenfrequency:
\[
L(t)=L_0\left(1+\varepsilon_L \sin\left[p\omega_1(1+\delta)t\right]
\right), \quad
u(t)=\varepsilon_u L_0\sin\left[p\omega_1(1+\delta)t +\varphi\right].
\]
Assuming $|\varepsilon_L|,|\varepsilon_u|\sim \varepsilon\ll 1$,
it is natural to look for
the solutions of equation (\ref{Qeq}) in the form similar to (\ref{ksi}),
\begin{equation}
Q_k^{(n)}(t)=\xi_k^{(n)}e^{-i\omega_k(1+\delta)t}
+\eta_k^{(n)}e^{i\omega_k(1+\delta)t},
\label{ksitime}
\end{equation}
but now we allow the
coefficients $\xi_k^{(n)}$ and $\eta_k^{(n)}$ to be
{\em slowly varying functions of time}.
The further procedure is well known in the theory of
parametrically excited systems [41-43].
%\cite{Louis,Land,Bogol}.
First we put
expression (\ref{ksitime}) into equation  (\ref{Qeq}) and neglect the terms
$\ddot{\xi },\ddot{\eta}$ (having in mind that $\dot{\xi },\dot{\eta}
\sim\varepsilon$, while $\ddot{\xi },\ddot{\eta}\sim\varepsilon^2$),
as well as the terms proportional to
$\dot{L}^2\sim\dot{u}^2\sim\varepsilon^2$.
Multiplying the resulting equation for $Q_k$ by the factors
$\exp\left[i\omega_k(1+\delta)t\right]$ and
$\exp\left[-i\omega_k(1+\delta)t\right]$ and performing
averaging over fast oscillations with the frequencies proportional to
$\omega_k$ (since the functions $\xi ,\eta$ practically do not change
their values at the time scale of $2\pi /\omega_k$)
one can verify that only the terms with the difference $j-k=\pm p$
survive in the right-hand side. Consequently, for {\it even\/} values of
$p$ the term $\dot{u}$ in $g_{kj}(t)$ does not make any contribution to
the simplified equations of motion, thus only the rate of change of the
cavity length $\dot{L}/L_0$ is important in this case.
On the contrary, if $p$ is an {\it odd\/} number, then the
field evolution depends on the velocity of the {\it centre of the cavity\/}
$v_c=\dot{u}+\dot{L}/2$ and does not depend on $\dot{L}$ alone.
These {\it interference effects\/} were discussed recently
(in the short time limit $\varepsilon\omega_1 t\ll 1$) in \cite{Ji98}
(see also \cite{Lamb}).
We assume hereafter that $u=0$ (i.e. that the left wall is at rest),
since this assumption does not change anything if $p$ is an even number,
whereas one should simply replace $\dot{L}/L_0$ by $2v_c/L_0$
if $p$ is an odd number.

The final equations for the coefficients
$\xi_k^{(n)}$ and $\eta_k^{(n)}$ contain only three terms with simple
{\it time independent\/} coefficients in the right-hand sides:
\begin{eqnarray}
\frac {\mbox{d}}{\mbox{d}\tau}\xi_k^{(n)}&=&
(-1)^p\left[(k+p)\xi_{k+p}^{(n)}- (k-p)\xi_{k-p}^{(n)}\right]
+2i\gamma k \xi_{k}^{(n)} ,
\label{pksik}\\
 \frac {\mbox{d}}{\mbox{d}\tau}\eta_k^{(n)}&=&
(-1)^p\left[(k+p)\eta_{k+p}^{(n)} -(k-p)\eta_{k-p}^{(n)}\right]
- 2i\gamma k \eta_{k}^{(n)}.
\label{petak}
\end{eqnarray}
The dimensionless parameters $\tau$ (a ``slow'' time) and $\gamma$ read
($\varepsilon\equiv\varepsilon_L$)
\begin{equation}
\tau =\frac 12\varepsilon\omega_1t, \qquad
\gamma=\delta/\varepsilon.
\label{tau}
\end{equation}
The initial conditions are
\begin{equation}\xi_k^{(n)}(0)=\delta_{kn},\qquad\eta_k^{(n)}(0)=0.
\label{ini}
\end{equation}
Note, however, that uncoupled equations (\ref{pksik})-(\ref{petak}) hold
only for $k\ge p$. This means that they describe the evolution of
{\it all\/} the Bogoliubov coefficients only if $p=1$. Then
{\it all\/} the functions $\eta_k^{(n)}(t)$ are {\it identically
equal to zero} due to the initial conditions (\ref{ini}),
consequently, no photon can be created from vacuum.
If $p\ge 2$, we have $p-1$ pair of {\it coupled\/} equations for the
coefficients with lower indices $1\le k\le p-1$
\begin{eqnarray}
\frac {\mbox{d}}{\mbox{d}\tau}\xi_k^{(n)}&=&
(-1)^p\left[(k+p)\xi_{k+p}^{(n)}- (p-k)\eta_{p-k}^{(n)}\right]
+2i\gamma k \xi_{k}^{(n)} ,
\label{pksikin}\\
 \frac {\mbox{d}}{\mbox{d}\tau}\eta_k^{(n)}&=&
(-1)^p\left[(k+p)\eta_{k+p}^{(n)} -(p-k)\xi_{p-k}^{(n)}\right]
- 2i\gamma k \eta_{k}^{(n)}.
\label{petakin}
\end{eqnarray}
In this case some functions $\eta_{k}^{(n)}(t)$ are not equal to zero at
$t>0$, thus we have the effect of photon creation from the vacuum.

It is convenient to introduce a new set of coefficients $\rho_k^{(n)}$,
whose lower indices run over all integers from $-\infty$ to $\infty$:
\begin{equation}
\rho_k^{(n)}=\left\{
\begin{array}{ll}
\xi_k^{(n)}\,, & k>0\\
0\,, & k=0\\
-\eta_{-k}^{(n)}\,, & k<0
\end{array}\right.
\label{defrho}
\end{equation}
Then one can verify that equations (\ref{pksik})-(\ref{petak}) and
(\ref{pksikin})-(\ref{petakin}) can be combined in a {\it single\/} set of
equation ($k=\pm 1, \pm 2, \ldots$)
\begin{equation}
\frac {\mbox{d}}{\mbox{d}\tau}\rho_k^{(n)}=
(-1)^p\left[(k+p)\rho_{k+p}^{(n)}- (k-p)\rho_{k-p}^{(n)}\right]
+2i\gamma k \rho_{k}^{(n)}
\label{prhok}
\end{equation}
with the initial conditions ($n=1,2,\ldots$)
\begin{equation}
\rho_k^{(n)}(0)=\delta_{kn}.
\label{inirho}
\end{equation}
A remarkable feature of the set of equations (\ref{prhok}) is that its
solutions satisfy {\it exactly\/}
the unitarity conditions (\ref{cond1})-(\ref{cond3})
(although the coefficients $\xi_k^{(n)}$ and $\eta_k^{(n)}$ introduced via
equation (\ref{ksitime}) have additional phase
factors in comparison with the coefficients defined in equation (\ref{ksi}),
these phases do not affect the identities concerned), which
can be rewritten as
\begin{eqnarray}
&& \sum_{m=-\infty}^{\infty}
m\rho_{m}^{(n)*}\rho_{m}^{(k)}
=n\delta_{nk}\,, \quad n,k=1,2,\ldots
\label{rhocond1} \\[2mm]
&&
 \sum_{n=1}^{\infty}\frac{m}{n}
\left[\rho_{m}^{(n)*}\rho_{j}^{(n)} -
\rho_{-m}^{(n)*}\rho_{-j}^{(n)} \right] =\delta_{mj}\,,
\quad m,j=1,2,\ldots
\label{rhocond2}\\[2mm]
&&
 \sum_{n=1}^{\infty}\frac{1}{n}
\left[\rho_{m}^{(n)*}\rho_{-j}^{(n)} -
\rho_{j}^{(n)*}\rho_{-m}^{(n)} \right] =0\,,
\quad m,j=1,2,\ldots
\label{rhocond3}
\end{eqnarray}
For example, calculating the derivative
$I=(d/d\tau)\sum_{m=-\infty}^{\infty}\, m\rho_{m}^{(n)*}\rho_{m}^{(k)}$
with the aid of equation (\ref{prhok}) and its complex conjugated
counterpart one can easily verify that $I=0$. Then the value of the
right-hand side of (\ref{rhocond1}) is a consequence of the initial
conditions (\ref{inirho}).
The identities (\ref{rhocond2}) and (\ref{rhocond3}) can be verified
in a similar way, if one uses instead of (\ref{prhok}) the recurrence
relations between the coefficients $\rho_{m}^{(n)}$ with the same lower
index $m$ but with different
{\it upper\/} indices derived in section~\ref{generic}.

Due to the initial conditions (\ref{inirho}) the solutions to (\ref{prhok})
satisfy the relation
\begin{equation}
\rho_{j+mp}^{(k+np)}\equiv 0 \quad {\rm if} j\neq k
\label{identrho}
\end{equation}
\[
j,k=0,1,\ldots,p-1, \quad m=0,\pm 1,\pm 2, \ldots\,,
\quad n=0, 1, 2, \ldots
\]
Consequently, the nonzero coefficients $\rho_m^{(n)}$ form
$p$ independent subsets
\begin{equation}
y_k^{(q,j)}\equiv\rho_{j+kp}^{(j+qp)}
\label{subsets}
\end{equation}
\[
j=0,1,\ldots,p-1, \quad q=0,1,2,\ldots\,,
\quad k=0,\pm 1,\pm 2, \ldots
\]
The subset $y_k^{(q,0)}$ is distinguished, because
$y_k^{(q,0)}\equiv 0$ for $k\le 0$ and the upper index $q$ begins at $q=1$.
This subset is considered in detail in section~\ref{semi}.
The generic case is studied in section~\ref{generic}.

\section{Total energy and the rate of photon generation}

It is remarkable that to
calculate the total energy of the field (normalized by $\hbar\omega_1$)
$${\cal E}(\tau)\equiv \sum_m m{\cal N}_m(\tau)$$
one does not need explicit expressions of the coefficients
$\rho_m^{(n)}(\tau)$.
Calculating the first and the second derivatives of ${\cal E}(\tau)$ with
the aid of the relations (\ref{defrho})-(\ref{rhocond3}) one can obtain
a simple differential equation (see \ref{Etot})
\begin{equation}
\ddot{\cal E}=4p^2a^2{\cal E} +4p^2\gamma^2{\cal E}(0) +
\frac{p^2}{6}(p^2-1)
+2p^2\gamma\sigma {\rm Im}({\cal G})
\label{eqEtot}
\end{equation}
where
\begin{equation}
 a= \sqrt{1-\gamma^2}\;, \quad \sigma=(-1)^p\,,
\label{def-a}
\end{equation}
\begin{equation}
{\cal G} = 2\sum_{n=1}^{\infty}
\sqrt{n(n+p)}\langle\hat {b}_n^{\dag}\hat {b}_{n+p}\rangle
+\sum_{n=1}^{p-1}\sqrt{n(p-n)}\langle\hat {b}_n\hat {b}_{p-n}\rangle
\label{defcalG}
\end{equation}
(if $p=1$, the last sum in (\ref{defcalG}) should be replaced by zero).
The quantum averaging is performed over the initial state of the field
(no matter pure or mixed).
The initial value of the total energy is
${\cal E}(0)= \sum_{n=1}^{\infty} n
\langle\hat {b}_n^{\dag}\hat {b}_{n}\rangle $,
whereas the initial value of the first derivative $\dot{\cal E}(\tau)$
reads (see \ref{Etot})
\begin{equation}
\dot{\cal E}(0)=-p\sigma{\rm Re}({\cal G})
\label{inconE}
\end{equation}
Consequently, the solution to equation (\ref{eqEtot}) can be expressed as
\begin{eqnarray}
{\cal E}(\tau)&=& {\cal E}(0) +\frac{2\sinh^2(pa\tau)}{a^2}
 \left[{\cal E}(0) +\frac{p^2-1}{24} +\frac{\gamma\sigma }{2}
 {\rm Im}({\cal G})\right] \nonumber\\
&-&\sigma{\rm Re}({\cal G}) \frac{\sinh(2pa\tau)}{2a}.
\label{ansEtot}
\end{eqnarray}
We see that the total energy increases exponentially at $\tau\to\infty$,
provided $\gamma< 1$.
In the special case $\gamma=0$ such asymptotical behaviour of the total
energy was obtained also in the frameworks of other approaches in
\cite{LawPRL,Law-new,Cole,Mep}. Here we have found the explicit
dependence of the total energy on time in the whole interval
$0\le \tau<\infty$, as well as a nontrivial dependence on the initial state
of field, which is contained in the constant parameter ${\cal G}$.
This parameter is equal to zero for initial Fock or
thermal states of the field. However, in a generic case ${\cal G}$ is
different from zero, and it can affect significantly the total energy,
if ${\cal E}(0)\gg 1$. Consider, for example, the case $p=2$.
If initially the first mode ($n=1$) was
in the coherent state $|\alpha\rangle$ with $\alpha=|\alpha|e^{i\phi}$,
$|\alpha|\gg 1$, and all other modes were not excited,
then ${\cal E}(0)=|\alpha|^2$, ${\cal G}=\alpha^2$, so
for $\tau\gg 1$ and $\gamma=0$ (exact resonance) we have
${\cal E}(\tau\gg 1)\approx \frac14 |\alpha|^2 e^{4\tau}
\left[2-\cos(2\phi)\right]$. The maximal value of the energy in this case
is three times bigger than the minimal one, depending on the phase $\phi$.

According to (\ref{ansEtot}), the initial stage of the evolution does not
depend on the detuning parameter $\gamma$ for all states which yield
Im$({\cal G})=0$, since at $\tau\to 0$ one has
\begin{equation}
{\cal E}(\tau) \approx {\cal E}(0)
-\sigma{\rm Re}({\cal G}) p\tau
+2\left[{\cal E}(0) +\frac{p^2-1}{24} +\frac{\gamma\sigma }{2}
 {\rm Im}({\cal G})\right] (p\tau)^2
\label{Etot-0}
\end{equation}
 Formula (\ref{Etot-0}) is {\it exact\/} in the case of $\gamma=1$.
If $\gamma>1$, then one should replace each function $\sinh(ax)/a$ in
(\ref{ansEtot}) by its trigonometrical counterpart
$\sin(\tilde{a}x)/\tilde{a}$, where
\begin{equation}
\tilde{a}= \sqrt{\gamma^2-1}
\label{def-tilda}
\end{equation}
In this case the total energy {\it oscillates\/} in time with
the period $\pi/(p\tilde{a})$, returning to the initial value at the
end of each period. For a large detuning $\gamma\gg 1$
the amplitude of oscillations decreases as $\gamma^{-1}$ if
Re${\cal G}\neq 0$ and as $\gamma^{-2}$ otherwise.
For the initial vacuum state of field we have
\begin{equation}
{\cal E}^{(vac)}(\tau)= \frac{p^2-1}{12a^2}\sinh^2(pa\tau)\,.
\label{Etotvac}
\end{equation}

The total number of photons in all the modes equals
${\cal N}={\cal N}^{(vac)}+{\cal N}^{(cav)}$, where
\begin{equation}
{\cal N}^{(vac)}=\sum_{m,n=1}^{\infty}\frac{m}{n}|\eta_m^{(n)}|^2
\label{Nvac}
\end{equation}
is the total number of photons generated from vacuum, and the sum
\begin{equation}
{\cal N}^{(cav)}= {\cal N}(0)
+ 2\sum_{m,n,k=1}^{\infty} \frac{m}{\sqrt{nk}}
\left[ \eta_{m}^{(n)*}\eta_{m}^{(k)}
\langle\hat {b}_n^{\dag}\hat {b}_k\rangle
+ {\rm Re}\left( \eta_{m}^{(n)}\xi_{m}^{(k)}
\langle\hat {b}_n\hat {b}_k \rangle\right)\right]
\label{Ncav}
\end{equation}
describes the influence of the initial state of the field
(to obtain the formula (\ref{Ncav}) one should take into account
the identity (\ref{cond1})).
Differentiating (\ref{Nvac}) and (\ref{Ncav}) with respect to $\tau$
and performing the summation over $m$ with the help of equations
(\ref{pksik})-(\ref{petakin}) or (\ref{prhok}) one can obtain the formulae
\begin{equation}
\frac {\mbox{d}{\cal N}^{(vac)}}{\mbox{d}\tau}=
2\sigma{\rm Re}\sum_{n=1}^{\infty}\frac 1n \sum_{m=1}^p m(p-m)
\rho_{-m}^{(n)*}(\tau )\rho_{p-m}^{(n)}(\tau )
\label{ratetot}
\end{equation}
\begin{eqnarray}
\frac{d{\cal N}^{(cav)}}{d\tau}&=& 2\sigma\sum_{n,k=1}^{\infty}
\frac{\langle\hat {b}_n^{\dag}\hat {b}_k\rangle }
{\sqrt{nk}} \sum_{m=1}^p m(p-m)
\left[ \rho_{-m}^{(n)*}\rho_{p-m}^{(k)}+
\rho_{-m}^{(k)}\rho_{p-m}^{(n)*}\right]\nonumber\\
&-& 2\sigma{\rm Re}\sum_{n,k=1}^{\infty}
\frac{\langle\hat{b}_n\hat{b}_k\rangle}
{\sqrt{nk}} \sum_{m=1}^p m(p-m)
\left[ \rho_{-m}^{(n)}\rho_{m-p}^{(k)}+
\rho_{m}^{(n)}\rho_{p-m}^{(k)}\right].
\label{extradot}
\end{eqnarray}
Consequently, to calculate the total number of photons one has to know
the coefficients $\eta_{m}^{(n)}$ and $\xi_{m}^{(n)}$ with the
lower indices $m=1,2,\ldots,p-1$.

\section{``Semi-resonance'' case ($p=1$)}\label{semi}

Let us start calculating the Bogoliubov coefficients with the
``semi-resonance'' case $p=1$. It is distinguished,
since all the coefficients $\eta_k^{(n)}(t)$ are equal to zero, and
the total number of photons is conserved.
In this specific case
one has to solve the set of equations ($k,n=1,2,\ldots$)
\begin{equation}
\frac {\mbox{d}}{\mbox{d}\tau}\xi_k^{(n)}=
(k-1)\xi_{k-1}^{(n)} - (k+1)\xi_{k+1}^{(n)} +2i\gamma k \xi_{k}^{(n)}
\label{1ksik}
\end{equation}
with the initial conditions $\xi_k^{(n)}(0)=\delta_{kn}$.
To get rid of the infinite number of equations we
introduce the {\em generating function}
\begin{equation}
X^{(n)}(z,\tau)=\sum_{k=1}^{\infty}\xi_k^{(n)}(\tau )z^k
\label{1defX}
\end{equation}
where $z$ is an auxiliary variable.
Using the relation $kz^k=z(\mbox{d}z^k/\mbox{d}z)$ one obtains
the first-order partial differential equation
\begin{equation}
\frac{\partial X^{(n)}}{\partial\tau}=\left(z^2-1 +2i\gamma z\right)
\frac{\partial X^{(n)}}{\partial z}+\xi_1^{(n)}(\tau)
\label{eqG}
\end{equation}
whose solution satisfying the initial condition $X^{(n)}(0,z)=z^n$ reads
\begin{equation}
X^{(n)}(z,\tau)=\left[\frac{z g(\tau) -S(\tau)}
{g^*(\tau)- zS(\tau)}\right]^n
+\int_0^{\tau} \xi_1^{(n)}(x)\,\mbox{d}x
\label{solG}
\end{equation}
where
\begin{equation}
S(\tau)=\sinh(a\tau)/a\,, \quad g(\tau)= \cosh(a\tau) + i\gamma S(\tau)\,.
\label{def-gS}
\end{equation}
Differentiating (\ref{solG}) over $z$ we find
\begin{equation}
\xi_1^{(n)}(\tau)=\frac{n[-S(\tau)]^{n-1}}
{[g^*(\tau)]^{n+1}}.
\label{sol-1n}
\end{equation}
Putting this expression into the integral in the right-hand side of
equation (\ref{solG}) we arrive at the final form of the generating function
\begin{equation}
X^{(n)}(z,\tau)=\left[\frac{z g(\tau) -S(\tau)}
{g^*(\tau)- zS(\tau)}\right]^n
-\left[\frac{ -S(\tau)}{g^*(\tau)}\right]^n
\label{solGfin}
\end{equation}
which satisfies automatically the necessary boundary condition
$X^{(n)}(\tau,0)=0$.
The right-hand side of (\ref{solGfin}) can be expanded into the power
series of $z$
with the aid of the formula (\cite{Bateman}, vol. 3, section 19.6,
equation (16))
\[
(1-t)^{b-c}(1-t+xt)^{-b}=\sum_{m=0}^\infty \frac{t^m}{m!} (c)_m
F(-m,b;c;x),
\]
where $F(a,b;c;x)$ means the Gauss hypergeometric function, and
$(c)_k\equiv \Gamma(c+k)/\Gamma(c)$.
In turn, the function $(c)_m F(-m,b;c;x)$  with an integer $m$
is reduced to the Jacobi polynomial in accordance with the formula
(\cite{Bateman}, vol. 2, section 10.8, equation (16))
\[
(c)_m F(-m,b;c;x)=m!(-1)^m P_m^{(b-m-c,\,c-1)}(2x-1).
\]
Consequently,
\begin{equation}
(1-t)^{b-c}(1-t+xt)^{-b}=\sum_{m=0}^\infty
(-t)^m P_m^{(b-m-c,\,c-1)}(2x-1)
\label{genBatJac}
\end{equation}
and the coefficient $\xi_m^{(n)}(\tau)$ reads
\begin{equation}
\xi_m^{(n)}(\tau)=(-\kappa)^{n-m}\lambda^{n+m}
P_m^{(n-m,\,-1)}\left(1-2\kappa^2\right)
\label{sol-mnJac}
\end{equation}
where
\begin{eqnarray}
\kappa(\tau)&=&\frac{S}{\sqrt{gg^*}}
\equiv \frac{S(\tau)}
{\sqrt{1+S^2(\tau)}}
\label{def-kap}\\
\lambda(\tau)&=&\sqrt{g(\tau)/g^*(\tau)}\equiv
\sqrt{1-\gamma^2\kappa^2}+i\gamma\kappa, \quad |\lambda|=1.
\label{def-lam}
\end{eqnarray}
The form (\ref{sol-mnJac}) is useful for $n\ge m$. To find a convenient
formula in the case of $n\le m$ we introduce the {\it two-dimensional\/}
generating function
\begin{eqnarray}
&&X(\tau,z,y)=\sum_{m=1}^\infty\sum_{n=1}^\infty z^m y^n
\xi_m^{(n)}(\tau)=\sum_{n=1}^\infty X^{(n)}(z,\tau)y^n \nonumber\\
&&=
\frac{ yz}{[g^*(\tau)+yS(\tau)]
[g^*(\tau) -g(\tau)yz+ S(\tau)(y-z)]}.
\label{G}
\end{eqnarray}
The coefficient at $z^m$ in (\ref{G}) yields another
one-dimensional generating function
\begin{equation} X_{m}(\tau,y)=\sum_{n=1}^\infty y^n \xi_m^{(n)}(\tau)
= y\frac{[g(\tau)y+S(\tau)]^{m-1}}
{[g^*(\tau) +yS(\tau)]^{m+1}}.
\label{Gm}\end{equation}
Then equation (\ref{genBatJac}) results in the expression
\begin{equation}
\xi_m^{(n)}=
(1-\kappa^2)\kappa^{m-n}\lambda^{n+m}
P_{n-1}^{(m-n,\,1)} \left(1-2\kappa^2\right).
\label{sol-nmJac}
\end{equation}
Note that the functions $S(\tau)$, $\cosh(a\tau)$ and $\kappa(\tau)$ are real
for any value of $\gamma$. For $\gamma>1$ it is convenient to use instead
of (\ref{def-gS}) the equivalent expressions in terms of the
trigonometrical functions:
\begin{equation}
\tilde{S}(\tau)=\sin(\tilde{a}\tau)/\tilde{a}\,, \quad
\tilde{g}(\tau)= \cos(\tilde{a}\tau) + i\gamma \tilde{S}(\tau)\,.
\label{def-tilgS}
\end{equation}
In the special case $\gamma=1$ one has
$S(\tau)=\tau$ and $g(\tau)=1 + i\tau$. In particular,
\begin{equation}
\xi_m^{(n)}(\tau;\gamma=1)
= \frac{\tau^{m-n}(1+i\tau)^{n-1}}
{(1-i\tau)^{m+1}}
P_{n-1}^{(m-n,\,1)}
\left(\frac{1-\tau^2}{1+\tau^2}\right).
\label{1a0}
\end{equation}

The knowledge of the two-dimensional generating function enables to
verify the unitarity condition (\ref{cond2}). Consider
the product $X^*(\tau,z_1,y_1)X(\tau,z_2,y_2)$, which is a four--variable
generating function for the products $\xi_m^{(n)*}\xi_l^{(k)}$.
Taking $y_1=\sqrt{u}\exp(i\varphi)$, $y_2^*=\sqrt{u}\exp(-i\varphi)$ and
integrating over $\varphi$ from $0$ to $2\pi$ one obtains
a three--variable generating function
$\sum z_1^{*m} z_2^l u^n \xi_m^{(n)*}\xi_l^{(n)}$.
Dividing it by $u$ and
integrating the ratio over $u$ from $0$ to $1$ one arrives finally at
the relation
\begin{equation}
\sum_{n,m,l=1}^{\infty} z_1^{*m} z_2^l \frac1n \xi_m^{(n)*}
\xi_l^{(n)}=-\ln\left(1-z_1^* z_2\right)=\sum_{k=1}^{\infty} \frac1k
\left(z_1^* z_2\right)^k,
\label{ident-2}
\end{equation}
which is equivalent to the special case of (\ref{cond2}) for
$\eta_{m}^{(k)}\equiv 0$:
\begin{equation}
\sum_{n}\; \frac1n \xi_{m}^{(n)*}(\tau)\xi_{j}^{(n)}(\tau)
\equiv \frac1m \delta_{mj}.
\label{cond2-1}
\end{equation}

Suppose that initially there was a single excited mode labeled with an
index $n$. Due to the linearity of the process one may assume that the mean
number of photons in this mode was $\nu_n=1$.
 Then the mean occupation number of the $m$-th mode at $\tau>0$ equals
\begin{equation}
{\cal N}_m^{(n)}=\frac{m}{n}\left[\xi_m^{(n)}\right]^2
= \frac{m}{n}\left[(1-\kappa^2) \kappa^{m-n}
P_{n-1}^{(m-n,\,1)}\left( 1-2\kappa^2 \right)\right]^2
\label{num-nm}
\end{equation}
where $\kappa$ is given by (\ref{def-kap}).
Although formula  (\ref{num-nm}) seems asymmetric with respect to the indices
$m$ and $n$, actually the relation
\begin{equation}
{\cal N}_m^{(n)}={\cal N}_n^{(m)}
\label{nm}
\end{equation}
holds. To prove it we calculate the generating function
\begin{equation}
Q(u,v)\equiv \sum_{m,n=1}^{\infty}v^m u^n {\cal N}_m^{(n)}.
\label{defgen-N}
\end{equation}
It is related to the function $X(z,y)$ (\ref{G}) as follows
\[
Q(u,v)= v\frac{d}{dv}\int_0^{u}dr\int_0^{2\pi}\int_0^{2\pi}
\frac{d\varphi d\psi}{(2\pi)^2}
X\left(\sqrt{r} e^{i\varphi},\sqrt{v} e^{i\psi}\right)
X^*\left(\sqrt{r} e^{i\varphi},\sqrt{v} e^{i\psi}\right).
\]
Having performed all the calculations we arrive at the expression
\begin{equation} 2Q(u,v)=
\frac{1+uv -\kappa^2(u+v)}
{\left\{ \left[1+uv -\kappa^2(u+v)\right]^2 -4uv(1-\kappa^2)^2\right\}^{1/2}}  -1.
\label{gen-N}
\end{equation}
Then (\ref{nm}) is a consequence of the relation $Q(u,v)=Q(v,u)$.

The initial stage of the evolution of
${\cal N}_m^{(n)}(\tau)$ does not depend on the detuning parameter $\gamma$,
since the principal term of the expansion of (\ref{num-nm}) with respect to
$\tau$ yields
\[
{\cal N}_{n\pm q}^{(n)}(\tau\to 0)= \frac{n\pm q}{n}
\left[\frac{n(n\pm 1)\ldots(n\pm q \mp 1)}{q!}\right]^2\tau^{2q}.
\]
However, the further evolution is sensitive to the value of $\gamma$.
If $\gamma\le 1$, then the function ${\cal N}_m^{(n)}(\tau)$ has
many maxima and minima (especially for large values of
$m$ and $n$), but finally it decreases asymptotically as
$mna^4/\cosh^4(a\tau)$. On the contrary, if $\gamma>1$, then
the function ${\cal N}_m^{(n)}(\tau)$ is periodic with the period
$\pi/\tilde{a}$, and it turns into zero for $\tau=k\pi/\tilde{a}$,
$k=1,2,\ldots$ (excepting the case $m=n$). The magnitude of
the coefficient ${\cal N}_m^{(n)}(\tau)$  decreases approximately
as $\gamma^{-2|m-n|}$ for $\gamma\gg 1$.

In the special case of a cavity filled in with a {\it high-temperature
thermal radiation\/},
the initial distribution over modes reads $\nu_n(T)=T/n$,
constant $T$ being proportional to the temperature. Then
${\cal N}_m^{\{T\}}=\sum_{n}\nu_n(T) {\cal N}_m^{(n)}$.
This sum is nothing but $T$ multiplied by the
coefficient at $v^m$ in the Taylor expansion of the function
\[ \tilde{Q}(v)=\int_0^1\frac{\mbox{d}u}{u}Q(u,v)=
\ln\frac{1-v\kappa^2(\tau)}{1-v}.\]
Thus we have
\[
{\cal E}_m^{\{T\}}=m{\cal N}_m^{\{T\}}=
T\left(1-[\kappa(\tau)]^{2m}\right).
\]
We see that the resonance vibrations of
the wall cause an effective cooling of the lowest electromagnetic modes
(provided $|\gamma|<1$).
The total number of quanta and the total energy in this example are
formally infinite,
due to the equipartition law of the classical statistical mechanics. In
reality both these quantities are finite, since
$\nu_n(T) < T/n$ at $n\to\infty$
due to the quantum corrections.
Other initial conditions in the special case of the
{\it exact\/} resonance ($\gamma=0$) were considered in \cite{D96}.
The total energy depends on time according to equation (\ref{ansEtot}) with
$p=1$.
An infinite growth of the energy of a classical string whose ends oscillate
at the frequency close to $\omega_1$ in the case of finite amplitude and
detuning ($\varepsilon\sim\delta\sim {\cal O}(1)$) was considered in
\cite{Dit}.

\section{Generic resonance case $p\ge 2$}
\label{generic}

Now we turn to calculating the nonzero Bogoliubov coefficients
$y_m^{(n,j)}(\tau)$ (\ref{subsets}) in the generic case $p\ge 2$.
One can easily verify that in the distinguished case $j=0$
the functions $y_m^{(n,0)}(\tau)$
 with $m\ge 1$ are given by the formulae for
$\xi_m^{(n)}(\tau)$ found in the preceding section, provided
one replaces $\tau$ by $\sigma p\tau$ and $\gamma$ by $\sigma\gamma$
(remember that $\sigma\equiv(-1)^p$),
whereas $y_m^{(n,0)}(\tau)\equiv 0$ for $m\le 0$.
In the generic case $j\neq 0$
it is reasonable to introduce a
generating function in the form of the {\it Laurent series\/} of an
auxiliary variable $z$
\begin{equation}
R^{(n,j)}(z,\tau)=\sum_{m=-\infty}^{\infty}y_m^{(n,j)}(\tau)z^m
\label{defR}
\end{equation}
since the lower index of the coefficient $y_m^{(n,j)}$ runs over all
integers from $-\infty$ to $\infty$.
One can verify that the function (\ref{defR}) satisfies the {\it homogeneous\/}
equation
\begin{equation}
\frac{\partial R^{(n,j)}}{\partial\tau}=\left[\sigma\left(
\frac1z -z\right) +2i\gamma \right]
\left(j+pz\frac{\partial }{\partial z}\right)R^{(n,j)}.
\label{eqR}
\end{equation}
The solution to (\ref{eqR}) satisfying
the initial condition $R^{(n,j)}(z,0)=z^n$ reads
\begin{equation}
R^{(n,j)}(z,\tau)=z^{-j/p}\left[\frac{z g(p\tau)
+\sigma S(p\tau)}{g^*(p\tau)+ z\sigma S(p\tau)}\right]^{n+j/p}
\label{solR}
\end{equation}
where the functions
$S(\tau)$ and $g(\tau)$ were defined in (\ref{def-gS}).
The coefficients of the Laurent series (\ref{defR}) can be calculated with
the aid of the Cauchy formula
\begin{equation}
y_m^{(n,j)}(\tau)=\frac{1}{2\pi i}\oint_{\cal C}\frac{dz}{z^{m+1}}
R^{(n,j)}(z,\tau)
\label{Cauchy}
\end{equation}
where the closed curve ${\cal C}$ rounds the point $z=0$ in the complex
plane in the counterclockwise direction. Making a scale transformation one
can reduce the integral (\ref{Cauchy}) with the integrand (\ref{solR})
to the integral representation of the Gauss hypergeometric function
(\cite{Bateman}, vol~1, section~2.1.3)
\begin{equation}
F(a,b;c;x)=\frac{-i\Gamma(c)\exp(-i\pi b)}{2\sin(\pi b)\Gamma(c-b)\Gamma(b)}
\oint_{1}^{(0+)}\frac{t^{b-1}(1-t)^{c-b-1}}{(1-tx)^a}dt,
\label{intpred}
\end{equation}
where ${\rm Re}(c-b)>0$, $b\neq 1,2,3,\ldots$, and the integration contour
begins at the point $t=1$ and passes around the point $t=0$ in the positive
direction. After some algebra one can obtain the expression
\begin{eqnarray}
y_m^{(n,j)}&=&
-\,\frac{\Gamma\left(-m-j/p\right)\Gamma\left(1+n+j/p\right)
\sin\left[\pi\left(m+j/p\right)\right]}
{\pi\Gamma\left(1+n-m\right) }
\nonumber\\
&\times&(\sigma\kappa)^{n-m}\lambda^{m+n+2j/p}
F\left(n+j/p\,,\,-m -j/p\,;\, 1+n-m\,;\, \kappa^2\right).
\label{solrhogen}
\end{eqnarray}
We assume hereafter $\kappa\equiv\kappa(p\tau)$ and
$\lambda\equiv\lambda(p\tau)$, the functions $\kappa(x)$ and
$\lambda(x)$ being defined as in (\ref{def-kap}) and (\ref{def-lam}).
Using the known formula
\begin{equation}
\Gamma(-z)\sin(\pi z)=-\pi/\Gamma(z+1)
\label{gammapm}
\end{equation}
one can eliminate the gamma-function of a negative argument:
\begin{eqnarray}
y_m^{(n,j)}&=&
\frac{\Gamma\left(1+n+j/p\right)
(\sigma\kappa)^{n-m}\lambda^{m+n+2j/p}}
{\Gamma\left(1+m+j/p\right)\Gamma\left(1+n-m\right) }
\nonumber\\
&\times&
F\left(n+j/p\,,\,-m -j/p\,;\, 1+n-m\,;\, \kappa^2\right).
\label{solrhogen1}
\end{eqnarray}
The form (\ref{solrhogen1}) gives an explicit expression for the
coefficient $\xi_{j+pm}^{(j+pn)}$ with $0\le m\le n$.
Moreover, it clearly shows the fulfilment of the initial
condition $y_m^{(n,j)}(\tau=0)=\delta_{mn}$.
Transforming the hypergeometric
function with the aid of the formula \cite{Bateman,Abram}
\[
\lim_{c\to -n}\frac{F(a,b;c;x)}{\Gamma(c)}=
\frac{(a)_{n+1}(b)_{n+1}x^{n+1}}{(n+1)!}
 F(a+n+1,b+n+1;n+2;x)
\]
($n=0,1,2,\ldots$) and the identity (\ref{gammapm})
one obtains an equivalent expression
\begin{eqnarray}
y_m^{(n,j)}&=&
\frac{ \Gamma\left(m+j/p\right)
(-\sigma\kappa)^{m-n}\lambda^{m+n+2j/p}}
{\Gamma\left(n+j/p\right) \Gamma\left(1+m-n\right) }
\nonumber\\
&\times&
F\left(m+j/p\,,\,-n -j/p\,;\, 1+m-n\,;\, \kappa^2\right)
\label{solrhogen2}
\end{eqnarray}
which gives a convenient form of the
coefficient $\xi_{j+pm}^{(j+pn)}$ for $m\ge n$.
Formula (\ref{solrhogen}) with negative values of the lower index gives an
explicit expression for the nonzero coefficients $\eta_{pk-j}^{(pn+j)}$
($k\ge 1,n\ge 0$):
\begin{eqnarray}
\eta_{pk-j}^{(pn+j)}&=&
-\,\frac{\Gamma\left(k-j/p\right)\Gamma\left(1+n+j/p\right)
\sin\left[\pi\left(k-j/p\right)\right]}
{\pi\Gamma\left(1+n+k\right) }
\nonumber\\
&\times&(\sigma\kappa)^{n+k}\lambda^{n-k+2j/p}
F\left(n+j/p\,,\,k -j/p\,;\, 1+n+k\,;\, \kappa^2\right).
\label{solrhogen3}
\end{eqnarray}
 Note that the expressions (\ref{solrhogen1})-(\ref{solrhogen3}) are valid
for $j=0$, too. In this case they coincide with the formulae obtained in
the preceding section. The formulae (\ref{solrhogen1})-(\ref{solrhogen3})
immediately give the short-time behaviour of the Bogoliubov coefficients
at $\tau\to 0$: it is sufficient to put $\kappa\approx  p\tau$,
$\lambda\approx 1$ and to replace the hypergeometric functions by $1$.
In this limit the detuning parameter $\gamma$ drops out of the expressions
(in the leading terms of the Taylor expansions).

At $\tau\to\infty$ we have the following asymptotics of the functions
$\kappa(p\tau)$ and $\lambda(p\tau)$ (if $\gamma\le 1$)
\[
\kappa\approx 1-\frac12 S^{-2}(p\tau) \to 1, \quad
\lambda\to a+i\gamma, \quad \tau\to\infty .
\]
Then
equation (\ref{solrhogen}) together with the known asymptotics of the
hypergeometric function $F(a,b;a+b+1;1-x)$ at $x\ll 1$
\cite{Bateman,Abram}
\begin{equation}
F(a,b;a+b+1;1-x)=\frac{\Gamma(a+b+1)}{\Gamma(a+1)\Gamma(b+1)}
\left[1+abx\ln(x) +{\cal O}(x)\right]
\label{F1}
\end{equation}
lead to the asymptotical expression for the Bogoliubov coefficients
\begin{eqnarray}
y_m^{(n,j)}(\tau\gg 1)&=& \frac{\sin[\pi(m+j/p)]}{\pi(m+j/p)}
(a+i\gamma)^{m+n+2j/p}\sigma^{n-m} \nonumber\\
&\times&\left[ 1+{\cal O}\left( \frac{mn}{S^2}\ln S\right)\right]
\label{asxieta}
\end{eqnarray}
For $\gamma<1$ the correction has an order $mn\tau\exp(-2ap\tau)$, while for
$\gamma=1$ it has an order $mn\ln(\tau)/\tau^2$.

One can verify that the generating function (\ref{solR})
satisfies the recurrence relation
\begin{equation}
\frac{\partial R^{(q,j)}}{\partial\tau} =(j+qp)\left\{\sigma\left[
R^{(q-1,j)} -R^{(q+1,j)}\right] +2i\gamma R^{(q,j)}\right\}
\label{recR}
\end{equation}
Its immediate consequence is an analogous relation for the
Bogoliubov coefficients with the same lower indices:
\begin{equation}
\frac{d }{d\tau}\rho_m^{(n)} = n\left\{\sigma\left[
\rho_m^{(n-p)} -\rho_m^{(n+p)}\right] +2i\gamma \rho_m^{(n)}\right\}.
\label{recrho}
\end{equation}
Equation (\ref{recrho}) is valid for $n>p$ (when
$q\ge 1$ and $j\ge 1$ in (\ref{recR})), since the coefficients
$\rho_m^{(n)}$ are not defined when $n< 0$.
However, using the chain of identities
\begin{eqnarray*}
&&R^{(-1,j)}(z)=
z^{-j/p}\left[\frac{S+gz}{g^*+ Sz}\right]^{j/p-1}
=\frac1z\left(\frac1z\right)^{j/p-1}\left[\frac{S +g^*/z}{g+ S/z}\right]
^{1-j/p} \\
&&= \frac1z\left[R^{(0,p-j)}(1/z^*)\right]^*
=\frac1z \sum_{k=-\infty}^{\infty} y_k^{(0,p-j)*}\left(\frac1z\right)^k
=\sum_{k=-\infty}^{\infty} y_{-k-1}^{(0,p-j)*}z^k
\end{eqnarray*}
one can obtain the first $p-1$ recurrence relations
\begin{equation}
\frac{d }{d\tau}\rho_m^{(n)} = n\left\{\sigma\left[
\rho_{-m}^{(p-n)*} -\rho_m^{(p+n)}\right] +2i\gamma \rho_m^{(n)}\right\},
\quad n=1,2,\ldots,p-1.
\label{recrho1}
\end{equation}
To treat the special case $n=p$ (it corresponds to the distinguished
subset with $j=0$) one should take
into account that $R^{(0,0)}(z)\equiv 1$, which means formally that
$\rho_m^{(0)}=\delta_{m0}$.
So the last recurrence relation reads
\[
\frac{d }{d\tau}\rho_m^{(p)} = p\left\{-\sigma\rho_m^{(2p)}
+2i\gamma \rho_m^{(p)}\right\}, \quad m\ge 1
\]
(remember that $\rho_m^{(p)}\equiv 0$ for $m\le 0$).
Now one can verify that the unitarity conditions
(\ref{rhocond2})-(\ref{rhocond3})
are the consequencies of the equations (\ref{recrho}) and (\ref{recrho1}).

Differentiating the ``vacuum'' part of sum (\ref{number}) with respect to
$\tau$ and performing
the summation over the upper index $n$ with the aid of
(\ref{recrho})-(\ref{recrho1}) (remembering that the
coefficients $\rho_m^{(n)}$
are different from zero provided the difference $n-m$ is a multiple of $p$)
one can obtain the formula for the photon generation rate from vacuum in
each mode ($0\le j\le p-1$, $q=0,1,2,\ldots$)
\begin{eqnarray}
&&\frac{d}{d\tau}{\cal N}_{j+pq}^{(vac)}=
-2\sigma(j+pq){\rm Re}\left[\xi_{j+pq}^{(j)}\eta_{j+pq}^{(p-j)}\right]
\nonumber\\[2mm]
&&=2 p\sqrt{1-\gamma^2\kappa^2}\,\frac{\sin(\pi j/p)\Gamma(q+j/p)
\Gamma(1+q+j/p)\Gamma(2-j/p)}{\pi \Gamma(j/p)\Gamma(q+1)\Gamma(q+2)}
\kappa^{2q+1}
\nonumber\\[2mm]
&&\times  F\left(q+j/p\,,\,-j/p\,;\,1+q\,;\,\kappa^2\right)
F\left(q+j/p\,,\,1-j/p\,;\,2+q\,;\,\kappa^2\right)
\label{ratejps}
\end{eqnarray}
We see that there is no photon creation in the modes with numbers
$p,2p,\ldots$.
At $\tau\ll 1$ we have $\dot{\cal N}_{j+pq}^{(vac)}\sim \tau^{2q+1}$. In the
long-time limit the photon generation rate tends to the constant value
(if $\gamma<1$)
\begin{equation}
\frac{d}{d\tau}{\cal N}_{j+pq}^{(vac)}=
\frac{2ap^2 \sin^2(\pi j/p)}{\pi^2 (j+pq)}
\left[1+{\cal O}\left(\frac{pq}{S^2}\ln S\right)\right], \quad ap\tau\gg 1
\label{asrate}
\end{equation}
For $q\gg 1$ and for a fixed value of $\kappa$ one can simplify the
right-hand side of (\ref{ratejps}) using Stirling's formula for the
Gamma-functions and the easily verified asymptotical formula
\[
F(a,b;c;z)\approx (1-az/c)^{-b}, \quad a,c\gg 1\, .
\]
In this case
\begin{equation}
\frac{d}{d\tau}{\cal N}_{j+pq}^{(vac)}\approx
2 p \sqrt{1-\gamma^2\kappa^2}\,
\frac{\sin(\pi j/p)\Gamma(2-j/p)\kappa^{2q+1}}
{\pi\Gamma(j/p)q^{2(1-j/p)}\left(1-\kappa^2\right)^{1-2j/p}},
\quad q\gg 1.
\label{asratebigq}
\end{equation}
In particular, if $q\gg S^2(p\tau)|\gg 1$, then
\begin{equation}
\frac{d}{d\tau}{\cal N}_{j+pq}^{(vac)}\approx
2 pa \,
\frac{\sin(\pi j/p)\Gamma(2-j/p)\left(S^2/q\right)^{2(1-j/p)} }
{\pi\Gamma(j/p) S^{2}}
 \exp\left(-q/S^2\right).
\label{asratebigqS}
\end{equation}
Comparing (\ref{asrate}) and (\ref{asratebigqS}) one can conclude
 that the number of the effectively excited modes
(i.e. the modes with a time independent photon generation rate) increases
in time exponentially, approximately as $S^2(\tau)/\ln S(\tau)$.

Differentiating equation (\ref{ratetot} ) once again over $\tau$
one can perform the summation over the upper index $n$ with the aid of
equations (\ref{recrho})-(\ref{recrho1}) to obtain a closed expression
for the {\it second derivative\/} of the total number of ``vacuum'' photons
\begin{eqnarray}
&&\frac{d^2}{d\tau^2}{\cal N}^{(vac)}=
2{\rm Re}\sum_{m=1}^{p-1} m(p-m)\left[\xi_{m}^{(m)}\xi_{p-m}^{(p-m)}
+\eta_{m}^{(p-m)*}\eta_{p-m}^{(m)*}\right]
\nonumber\\[2mm]
&&=2 \sum_{m=1}^{p-1} m(p-m)\left\{m(p-m)\left[\frac{\kappa}{p}
F\left(\frac{m}{p}\,,\,1-\frac{m}{p}\,;\,2\,;\,\kappa^2\right)\right]^2
\right.
\nonumber\\[2mm]
&&\left.+ \left(1-2\gamma^2\kappa^2\right)
F\left(\frac{m}{p}\,,\,-\frac{m}{p}\,;\,1\,;\,\kappa^2\right)
F\left(\frac{m}{p}-1\,,\,1-\frac{m}{p}\,;\,1\,;\,\kappa^2\right)\right\}
\label{ratetot2}
\end{eqnarray}
In the short-time limit one obtains
\begin{equation}
\ddot{\cal N}^{(vac)}=\frac13 p(p^2-1), \quad |ap\tau|\ll 1
\label{2dersmall}
\end{equation}
In the long-time limit the formulae (\ref{gammapm}), (\ref{F1}) and
$\sum_{m=1}^{p-1} \sin^2(\pi m/p)=p/2$ lead to another simple expression
(provided $p\ge 2$)
\begin{equation}
\ddot{\cal N}^{(vac)}=2a^2p^3/\pi^2, \quad ap\tau\gg 1, \quad a>0
\label{2derbig}
\end{equation}
Consequently, the total number of photons created from vacuum due to NSCE
increases in time quadratically both in the short-time and in the long-time
limits (although with different coefficients).

It is interesting to compare formula (\ref{2derbig}) with the total rate
of change of the number of ``cavity'' photons due to
nonvacuum initial conditions. Using equation (\ref{extradot}) and
replacing the coefficients $\rho_m^{(n)}$ by their asymptotical values
(\ref{asxieta}) one can obtain the expression
\begin{eqnarray}
&&\frac{d{\cal N}^{(cav)}}{d\tau}= \frac{4ap^2}{\pi^2}
\sum_{m=1}^{p-1}\sin^2(\pi m/p) \sum_{n,k=0}^{\infty}
\frac{\sigma^{n+k}}{\sqrt{(m+pn)(m+pk)}}\nonumber\\[2mm]
&&\times \left\{
\langle\hat {b}_{m+pn}^{\dag}\hat {b}_{m+pk}\rangle ( a+i\gamma)^{k-n}
-\sigma{\rm Re}\left[\langle\hat{b}_{m+pn}\hat{b}_{m+pk}\rangle
( a+i\gamma)^{k+n+1} \right]\right\}
\label{extradotas}
\end{eqnarray}
which holds provided $ap\tau\gg 1$ and  $a>0$.
For the physical initial states the sum in the right-hand side
of (\ref{extradotas}) is finite. This is obvious if a finite
number of modes was excited initially. But even if the cavity was initially
in a high-temperature thermal state, so that
$\langle\hat {b}_{n}^{\dag}\hat {b}_{k}\rangle=\delta_{nk}T/n$,
$\langle\hat {b}_{n}\hat {b}_{k}\rangle=0$, the sum over $n,k$ yields a
finite value
$
T \sum_{n=0}^{\infty}\,(m+pn)^{-2}
$.
Consequently, the total number of ``nonvacuum''
photons increases in time {\it linearly\/} at $ap\tau\gg 1$, whereas the
total number of quanta generated from vacuum increases {\it quadratically\/}
in the long time limit.
At the same time, the total ``vacuum'' and ``nonvacuum'' energies increase
exponentially if $\gamma<1$ (see section~3).
The origin of the difference in the behaviours of the total energy and the
total number
of photons becomes clear, if one looks at the asymptotical formulae
(\ref{asrate})-(\ref{asratebigqS}). They show that the rate of photon
generation in the $m$th completely excited
mode decreases approximately as $1/m$ (excepting the modes whose numbers
are multiples of $p$), so the stationary rate of the {\it energy\/}
generation asymptotically almost does not depend on $m$.
In turn, the number of the effectively excited modes increases
in time exponentially. These two factors lead to
the exponential growth of the total energy
(see also \cite{Klim} in the special case $\gamma=0$).

\section{The ``principal resonance'' ($p=2$)}

Some formulae obtained in the preceding section can be
simplified in the special case $p=2$. In this case there are two subsets
of nonzero Bogoliubov coefficients.
The first one consists of the coefficients with even upper
and lower indices $\xi_{2k}^{(2q)}$ which are reduced to the coefficients
$\xi_{k}^{(q)}$ of the ``semi-resonance'' case
(since $\eta_{2k}^{(2q)}\equiv 0$,
this subset does not contribute to the generation of new photons).
The second subset is formed by the ``odd'' coefficients which can be
written as [$\kappa\equiv \kappa(2\tau)$]
\begin{eqnarray}
\xi_{2m+1}^{(2n+1)}&=&
\frac{\Gamma\left(n+3/2\right)
\kappa^{n-m}\lambda^{m+n+1}}
{\Gamma\left(m+3/2\right)\Gamma\left(1+n-m\right) }
\nonumber\\
&\times&
F\left(n+1/2\,,\,-m -1/2\,;\, 1+n-m\,;\, \kappa^2\right),
\quad n\ge m
\label{xinm}
\end{eqnarray}
\begin{eqnarray}
\xi_{2m+1}^{(2n+1)}&=&
\frac{(-1)^{m-n} \Gamma\left(m+1/2\right)
\kappa^{m-n}\lambda^{m+n+1}}
{\Gamma\left(n+1/2\right) \Gamma\left(1+m-n\right) }
\nonumber\\
&\times&
F\left(m+1/2\,,\,-n -1/2\,;\, 1+m-n\,;\, \kappa^2\right),
\quad m\ge n
\label{ximn}
\end{eqnarray}
\begin{eqnarray}
\eta_{2k+1}^{(2n+1)}&=&
\frac{(-1)^{k-1}\Gamma\left(k+1/2\right)\Gamma\left(n+3/2\right)
\kappa^{n+k+1}\lambda^{n-k} }
{\pi\Gamma\left(2+n+k\right) }
\nonumber\\
&\times&
F\left(n+1/2\,,\,k +1/2\,;\, 2+n+k\,;\, \kappa^2\right).
\label{etank}
\end{eqnarray}
All the ``odd'' coefficients can be expressed
in terms of the complete elliptic integrals \cite{BrMar}
\[
{\bf K}(\kappa )=\int_0^{\pi /2}\frac {\mbox{d}\alpha}{\sqrt {1
-\kappa^2\sin^2\alpha}}\, , \quad
{\bf E}(\kappa )=\int_0^{\pi /2}\mbox{d}\alpha
\sqrt {1-\kappa^2\sin^2\alpha}\,.
\]
In particular,
\begin{equation}
\xi_1^{(1)}=\frac{2}{\pi}\lambda(\kappa){\bf E}(\kappa), \quad
\eta_1^{(1)}=\frac{2}{\pi\kappa}\left[\tilde{\kappa}^2{\bf K}(\kappa )
-{\bf E}(\kappa )\right],
\label{xietell}
\end{equation}
where
\begin{equation}
\tilde{\kappa}\equiv\sqrt{1-\kappa^2}=
\left[1 +S^2(2\tau)\right]^{-1/2}.
\label{deftilkappa}
\end{equation}
However, the analogous expressions for the coefficients $\xi_m^{(n)}$ and
$\eta_m^{(n)}$ with $m,n>1$ appear rather cumbersome (they can be written
as linear combinations of the functions ${\bf E}(\kappa )$ and
${\bf K}(\kappa )$ multiplied by some rational functions of $\kappa$ and
$\tilde{\kappa}$), so we do not bring them here.

The photon generation rate from vacuum in the
principal cavity mode ($m=1$) reads
\begin{equation}
\frac {\mbox{d}{\cal N}_1^{(vac)}}{\mbox{d}\tau}=
-2{\rm Re}\left[\eta_1^{(1)} \xi_1^{(1)}\right]
=\frac {8\sqrt{1-\gamma^2\kappa^2}}{\pi^2\kappa}
{\bf E}(\kappa )\left[{\bf E}(\kappa )-
\tilde{\kappa}^2{\bf K}(\kappa )\right].
\label{rate1}
\end{equation}
The total number of photons in the first mode can be obtained by
integrating equation (\ref{rate1}). Taking into account the relation
\begin{equation}
\sqrt{1-\gamma^2\kappa^2}\mbox{d}\tau =\mbox{d}\kappa/\tilde{\kappa}^2
\label{dtau}
\end{equation}
and the differentiation rules for the complete elliptic integrals
\begin{equation}
\frac {\mbox{d}{\bf K}(\kappa )}{\mbox{d}\kappa}=\frac {
{\bf E}(\kappa )}{\kappa\tilde{\kappa}^2}-\frac {{\bf K}(\kappa )}{
\kappa},\quad
\frac {\mbox{d}{\bf E}(\kappa )}{\mbox{d}\kappa}=\frac {
{\bf E}(\kappa )-{\bf K}(\kappa )}{\kappa}
\label{difrul}
\end{equation}
one can verify the following result:
\begin{equation}{\cal N}_1^{(vac)}(\kappa )=
\frac 2{\pi^2}{\bf K}(\kappa )
\left[2{\bf E}(\kappa) -\tilde{\kappa}^2{\bf K}(\kappa )\right]
-\frac 12.
\label{num1EK}
\end{equation}
Making the transformation \cite{Bateman,Abram}
\[
{\bf K}\left(\frac{1-\tilde{\kappa}}{1+\tilde{\kappa}}\right)
=\frac{1+\tilde{\kappa}}{2}{\bf K}(\kappa), \quad
{\bf E}\left(\frac{1-\tilde{\kappa}}{1+\tilde{\kappa}}\right)
=\frac{{\bf E}(\kappa)+\tilde{\kappa}{\bf K}(\kappa)}
{1+\tilde{\kappa}}
\]
one can rewrite formulae (\ref{xietell}) and (\ref{num1EK}) in the form given
in \cite{DKPR} for $\gamma=0$.
Using the asymptotical expansions of the elliptic integrals at $\kappa\to 1$
\cite{Grad}
\begin{eqnarray*}
{\bf K}(\kappa )&\approx&\ln\frac 4{\tilde{\kappa}}
+\frac 14\left(\ln\frac 4{\tilde{\kappa}}-1\right)\tilde{\kappa}^
2+\cdots \\
{\bf E}(\kappa )&\approx& 1+\frac 12\left(\ln\frac
4{\tilde{\kappa}}-\frac 12\right)\tilde{\kappa}^2 +\cdots
\end{eqnarray*}
one can obtain the formula
\begin{equation}
{\cal N}_1^{(vac)}(\tau\gg 1)=\frac {8a}{\pi^2}\tau +
\frac4{\pi^2}\ln\left(\frac{2}{a}\right)-\frac 12 +
{\cal O}\left(\tau e^{-4a\tau}\right), \quad a>0.
\label{num1as}
\end{equation}
In the special case of $\gamma=1$ one can obtain the expansion
\[
{\cal N}_1^{(vac)}(\tau\gg 1)=\frac {4}{\pi^2}\ln\tau +
\frac{12}{\pi^2}\ln2-\frac 12 + {\cal O}\left(\tau^{-2}\right)
\]
If $\gamma>1$, the number of photons in the principal mode oscillates with
the period $\pi/(2\tilde{a})$. For $\gamma\gg 1$ one can write
$\kappa\approx\sin(2\tilde{a}\tau)/\tilde{a}$, i.e. $|\kappa|\ll 1$.
In this case
\[
{\cal N}_1^{(vac)}\approx \frac{\kappa^2}{4}\approx
\frac{\sin^2(2\tilde{a}\tau)}
{4\tilde{a}^2} \ll 1.
\]
The second derivative of the total number of ``vacuum'' photons can be
written as
\begin{eqnarray}
&&\frac {\mbox{d}^2{\cal N}^{(vac)}}{\mbox{d}\tau^2}=
2\left[{\rm Re}\left(\left[\xi_1^{(1)}\right]^2\right)+
\left|\eta_1^{(1)}\right|^2\right]
\nonumber\\
&&=
\frac 8{\pi^2\kappa^2}
\left[\tilde{\kappa}^4{\bf K}^2(\kappa )
-2\tilde{\kappa}^2{\bf K}(\kappa ){\bf E}(\kappa)
+\left(1+\kappa^2 -2\gamma^2\kappa^4\right){\bf E}^2(\kappa)\right]
\label{sectot}
\end{eqnarray}
In the limiting cases this formula yields
\[
{\cal N}^{(vac)}(\tau\ll 1)\approx\tau^2
\]
\[
{\cal N}^{(vac)}(\tau\gg 1)= 8a^2\tau^2/{\pi^2} +{\cal O}(\tau),
 \quad a>0.
 \]
If $\gamma\gg 1$, then $|\kappa|\ll 1$, but
$\gamma^2\kappa^2\approx\sin^2(2\tilde{a}\tau)\sim {\cal O}(1)$. In this
case the Taylor expansion of the expression (\ref{sectot}) yields
$\ddot{{\cal N}}^{(vac)}=2\cos(4\tilde{a}\tau) + {\cal O}(\gamma^{-2})$.
Integrating this equation with account of the initial conditions
$\dot{{\cal N}}^{(vac)}(0)={\cal N}^{(vac)}(0)=0$ one obtains
$ {\cal N}^{(vac)}\approx {\cal N}_1^{(vac)}\approx
\sin^2(2\tilde{a}\tau)/(4\tilde{a}^2)$.

\section{Discussion}

Let us discuss briefly the main results of the paper. We have solved the
problem of the photon generation due to the nonstationary Casimir effect in
an ideal Fabry-Perot cavity with an equidistant spectrum, if the cavity
walls perform small (quasi)resonance oscillations at the frequency
$\omega_w=p(\pi c/L_0)(1+\delta)$, for any integer value of $p=1,2,\ldots$.
Namely,
we have found explicit analytical expressions for the Bogoliubov coefficients,
the rate of photon production in each mode and the total energy in the case
of an arbitrary (although small compared with $\omega_w$) detuning.
These expressions are {\it exact\/} consequences of the reduced equations
(\ref{prhok}) or (\ref{recrho})-(\ref{recrho1}). One should
remember, however, that the reduced
equations arise after averaging the exact equation (\ref{Qeq}) over fast
oscillations and neglecting the second-order terms with respect to small
parameters $\varepsilon$ and $\delta$. Consequently, the ``true'' functions
${\cal N}(t)$, ${\cal E}(t)$, etc. could differ from those given above in
terms proportional to $\varepsilon^2$. But such a difference seems quite
insignificant under the realistic conditions. As was shown in
\cite{D95,DKPR}, it is hardly possible to obtain the value
of the dimensionless amplitude of the {\it resonance\/} wall vibrations
$\varepsilon$ exceeding $10^{-8}$ in a laboratory.
This means that the relative difference
between the ``true'' magnitude of the photon generation rate (for example)
and that given in section 5 could be of the order of $10^{-8}$ (or less)
for $t<t_c\sim (\omega_1\varepsilon^2)^{-1}$. For
$\omega_1\sim 10^{10}$ s$^{-1}$ the characteristic time $t_c$ has an order
of months or years, and even for the optical frequences it has an order of
seconds (although it is unclear how to cause the wall to vibrate at an
optical frequency with a sufficiently big amplitude).
        Another argument in favour of the solutions obtained
is that these solutions satisfy {\it exactly\/} the Bogoliubov
transformation unitarity conditions (\ref{cond1})-(\ref{cond3}).

Note that the rate of photon generation from vacuum in some mode is
proportional to $p^2\varepsilon$ (if $\gamma=0$), and the total
generation rate is proportional to $p^3\varepsilon^2$. Actually, the
dimensionless amplitude of the wall oscillations $\varepsilon$ is
inversly proportional to the frequency, since it is determined by the
maximal possible stresses inside the wall \cite{D95,DKPR}. Thus we see
that increasing the resonance frequency one could achieve, in principle,
some amplification of the number of photons proportional to $p$.

It was shown in the previous studies [10,14-23]
%\cite{DK92,DKN93,Law,LawPRL,Law-new,D95,Cole,Mep,DKPR,D96,Lamb}
that the photon
production from vacuum due to the NSCE {\it could\/} be observed under the
condition of the strict parametric resonance. Here it is demonstrated
explicitly that the photons {\it cannot\/} be produced if the detuning
$\delta$ exceeds the dimensionless amplitude $\varepsilon$. This result
confirms once again the statement made in \cite{DKPR} that the
NSCE could be observed only in the resonance regime, ruling out the
nonresonance laws of motion of the wall. The requirements to a possible
experiment turn out rather hard (for example, for the principal
frequency about $10$~GHz the detuning should not exceed $100$~Hz for
the time of the order of at least $0.01$~s), but they do not seem to be
absolutely unrealizable.

Another source of troubles is connected with
a nonideality of real cavities. Until now there were only few attempts
to take into account different losses in the cavities with moving
boundaries \cite{Lamb,D98,Lamb98}, and this problem is still a challenge for
theoreticians.

\newpage
\appendix

\section{}\label{Etot}

Using equations (\ref{number}) and (\ref{defrho}) one can express
the total energy in all the modes as
\[
{\cal E}=
\sum_{n=1}^{\infty} \frac1n S^{(n)}
+ \sum_{n,k=1}^{\infty} \frac{\langle\hat {b}_n^{\dag}\hat {b}_k\rangle }
{\sqrt{nk}}U_1^{(nk)}
+{\rm Re}\sum_{n,k=1}^{\infty} \frac{\langle\hat {b}_n\hat {b}_k\rangle }
{\sqrt{nk}}U_2^{(nk)}
\]
where
\begin{equation}
S^{(n)}=\sum_{m=1}^{\infty}m^2\left|\rho_{-m}^{(n)}\right|^2,
\label{defSn}
\end{equation}
\begin{equation}
U_1^{(nk)}=\sum_{m=-\infty}^{\infty}m^2\rho_{m}^{(n)*}
\rho_{m}^{(k)}, \quad
U_2^{(nk)}=-\sum_{m=-\infty}^{\infty}m^2\rho_{m}^{(n)}\rho_{-m}^{(k)}
\label{defU12}
\end{equation}
(to write $U_2^{(nk)}$ as a sum from $-\infty$ to $\infty$ one should take
into account that the summand in the last sum of (\ref{number}) is
symmetrical with respect to $n$ and $k$).
Differentiating $U_1^{(nk)}$ with respect to $\tau$ and taking into account
the equations (\ref{prhok}) one can obtain after a simple algebra the
expression
\begin{equation}
\frac{d}{d\tau}U_1^{(nk)}=-p(-1)^p
\sum_{m=-\infty}^{\infty} m(m+p)\left[
\rho_{m}^{(n)*} \rho_{m+p}^{(k)} +
\rho_{m}^{(k)}\rho_{m+p}^{(n)*}\right].
\label{dotU1}
\end{equation}
Differentiating the above expression once more one obtains
\[
\ddot{U}_1^{(nk)}=4p^2 U_1^{(nk)} + 2i\gamma p^2(-1)^p\chi_1^{(nk)},
\]
where
\[
\chi_1^{(nk)}=
\sum_{m=-\infty}^{\infty} m(m+p)\left[
\rho_{m}^{(k)}\rho_{m+p}^{(n)*}
-\rho_{m}^{(n)*} \rho_{m+p}^{(k)} \right].
\]
Differentiating $\chi_1^{(nk)}$ one can verify that
\[
\dot\chi_1^{(nk)}=2i\gamma(-1)^p\dot{U}_1^{(nk)}.
\]
Consequently,
\[
\chi_1^{(nk)}=2i\gamma(-1)^p U_1^{(nk)} +const,
\]
where the additive constant
is determined by the initial conditions. Finally we arrive at the equation
\[
\ddot{U}_1^{(nk)}=4p^2(1-\gamma^2)U_1^{(nk)} +
4p^2\gamma^2 U_1^{(nk)}(0) + 2i\gamma p^2(-1)^p \chi_1^{(nk)}(0)
\]
with
\[
U_1^{(nk)}(0)=n^2\delta_{nk}, \quad
\chi_1^{(nk)}(0)=nk\left[\delta_{k,n-p} -\delta_{n,k-p}\right].
\]
Using the same scheme one can obtain analogous relations for the
coefficient $U_2^{(nk)}$:
\begin{equation}
\frac{d}{d\tau}U_2^{(nk)}=
p(-1)^p \sum_{m=-\infty}^{\infty} m\rho_{m}^{(n)} \left[ (m+p)
\rho_{-m-p}^{(k)} -(p-m) \rho_{p-m}^{(k)}\right],
\label{dotU2}
\end{equation}
\[
\ddot{U}_2^{(nk)}=4p^2 U_2^{(nk)} - 2i\gamma p^2(-1)^p\chi_2^{(nk)},
\]
\[
\chi_2^{(nk)}=
\sum_{m=-\infty}^{\infty} m\rho_{m}^{(n)} \left[ (m+p)
\rho_{-m-p}^{(k)} +(p-m) \rho_{p-m}^{(k)}\right],
\]
\[
\dot\chi_2^{(nk)}=-2i\gamma(-1)^p\dot{U}_2^{(nk)},
\]
\[
\ddot{U}_2^{(nk)}=4p^2(1-\gamma^2)U_2^{(nk)}
- 2i\gamma p^2(-1)^p \chi_2^{(nk)}(0),
\]
\[
\chi_2^{(nk)}(0)=nk\delta_{k,p-n} .
\]
The calculation of the vacuum contribution to the total energy
\[
{\cal E}^{(vac)} =\sum_{n=1}^{\infty} \frac1n S^{(n)}
\]
is more involved, since the summation in (\ref{defSn}) is performed now not
from $-\infty$ to $\infty$, but over
the coefficients $\rho_{m}^{(n)}$ with {\it negative\/} indices $m$ only.
Differentiating the sum (\ref{defSn}) with respect to $\tau$
and using equations (\ref{prhok}) we obtain
\begin{equation}
\dot {S}^{(n)}= 2(-1)^p {\rm Re}
\sum_{m=1}^{\infty} m^2\rho_{-m}^{(n)} \left[ (m+p)
\rho_{-m-p}^{(n)*} +(p-m) \rho_{p-m}^{(n)*}
\right].
\label{dotSn}
\end{equation}
Differentiating the expression (\ref{dotSn}) once again one can obtain
after some algebra the equation
\begin{eqnarray}
\ddot {{\cal E}}^{(vac)}&=& 4p^2 {\cal E}^{(vac)}
+4p(-1)^p\gamma\sum_{n=1}^{\infty} \frac1n \Phi^{(n)}\nonumber\\
&+& 2{\rm Re}\sum_{m=1}^p m(p-m)^2 \left[ m F_m +(m+p)G_m\right],
\label{ddotsum}
\end{eqnarray}
where
\[
\Phi^{(n)}={\rm Im}
\sum_{m=1}^{\infty} m^2\rho_{-m}^{(n)} \left[
(p-m) \rho_{p-m}^{(n)*} -
(m+p)\rho_{-m-p}^{(n)*} \right]
\]
\[
F_m=
 \sum_{n=1}^{\infty}\frac{1}{n}
\left[\rho_{m}^{(n)*}\rho_{m}^{(n)} -
\rho_{-m}^{(n)*}\rho_{-m}^{(n)} \right]
\]
\[
G_m=
 \sum_{n=1}^{\infty}\frac{1}{n}
\left[\rho_{m+p}^{(n)*}\rho_{m-p}^{(n)} -
\rho_{p-m}^{(n)*}\rho_{-m-p}^{(n)} \right]
\]
Differentiating the function $\Phi^{(n)}$ over $\tau$ and using again
equations (\ref{prhok}) one can verify that the derivative
$d\Psi/d\tau$ of the combination
$
\Psi\equiv \sum_{n=1}^{\infty}\;\frac1n\left[\Phi^{(n)}
+p(-1)^p\gamma S^{(n)}\right]
$
can be written in the form analogous to the last sum (from $1$ to $p$)
of equation (\ref{ddotsum}), but the symbol Re should be replaced by Im.
Since $F_m=1/m$ due to the identity (\ref{rhocond2}) and
$G_m=0$ due to (\ref{rhocond3}), we have $d\Psi/d\tau=0$. Taking into
account the initial conditions $\Phi^{(n)}(0)=S^{(n)}(0)=0$ one obtains
$\Psi(\tau)= 0$. Combining all the terms giving the second derivative of
${\cal E}$ one can arrive finally at equation (\ref{eqEtot}), where the
term $\frac16 p^2(p^2-1)$ is the value of the sum
$2\sum_{m=1}^p \,m(p-m)^2$.

The initial value of the first derivative $\dot{\cal E}(\tau)$ is determined
by the right-hand sides of equations (\ref{dotU1}), (\ref{dotU2}) and
(\ref{dotSn}) taken at $\tau=0$, when $\rho_m^{(n)}=\delta_{mn}$:
\[
\dot{\cal E}(0) = -2p\sigma\sum_{n=1}^{\infty}
\sqrt{n(n+p)}{\rm Re}\langle\hat {b}_n^{\dag}\hat {b}_{n+p}\rangle
-p\sigma\sum_{n=1}^{p-1}\sqrt{n(p-n)}{\rm Re}
\langle\hat {b}_n\hat {b}_{p-n}\rangle  .
\]
Comparing this formula with (\ref{defcalG}) we arrive at equation
(\ref{inconE}).

\end{document}